\documentclass{article}
\usepackage{amsmath}
\usepackage{graphicx}

\newtheorem{theorem}{Theorem}[section]

\newtheorem{definition}[theorem]{Definition}
\newtheorem{example}[theorem]{Example}

\newtheorem{lemma}[theorem]{Lemma}
\newtheorem{notation}[theorem]{Notation}

\newtheorem{remark}[theorem]{Remark}

\newenvironment{proof}[1][Proof]{\textbf{#1.} }{\ \rule{0.5em}{0.5em}}
\input{tcilatex}

\begin{document}

\title{The Non-Anticipation of the Asynchronous Systems}
\author{Serban E. Vlad \\
Oradea City Hall, Piata Unirii, Nr. 1, 410100, Oradea, Romania\\
serban\_e\_vlad@yahoo.com}
\date{}
\maketitle

\begin{abstract}
The asynchronous systems are the models of the asynchronous circuits from
the digital electrical engineering and non-anticipation is one of the most
important properties in systems theory. Our present purpose is to introduce
several concepts of non-anticipation of the asynchronous systems.
\end{abstract}

\section{Introduction. Bibliography-Related Remarks}

The asynchronous systems are the mathematical models of the asynchronous
circuits from the digital electrical engineering and our challenge is the
construction of an asynchronous systems theory. The insufficient
bibliography that we have at disposal is probably influenced by the great
importance of the topic (researchers do not publish) and it consists of:

a) mathematical studies in switching theory from the 60's and we mention
here the name of Grigore Moisil that used the discrete time modeling of the
asynchronous circuits;

b) engineering studies, which are always non-formalized and dedicated to
applications. Such a literature creates intuition, but it does not give
acceptable models or tools of investigation;

c) mathematical literature that can produce analogies.

The study of the asynchronous systems is closely connected to the notion of
signal, meaning a 'nice' $\mathbf{R}\rightarrow \{0,1\}^{n}$ function. In
this context we mention the apparent total absence of the mathematicians'
interest in the study of the $\mathbf{R}\rightarrow \{0,1\}$ functions, that
should be an interesting direction of investigation in temporal logic too.

The paper is dedicated to non-anticipation, one of the most important
properties of the systems. We exemplify its use by showing how 0 can be
chosen as initial time, how the transfers of the systems are composed and
how the asynchronous real time non-deterministic systems behave in certain
circumstances in a synchronous discrete time deterministic way.

\section{Preliminaries}

\begin{definition}
$\mathbf{B}=\{0,1\}$ endowed with the laws $\overline{\;\;},\cup ,\cdot
,\oplus $ is called the \textbf{binary Boole algebra}.
\end{definition}

\begin{notation}
$Seq=\{(t_{k})|t_{k}\in \mathbf{R},k\in \mathbf{N},t_{0}<t_{1}<t_{2}<...\;$%
unbounded from above$\}.$
\end{notation}

\begin{notation}
The restriction of the function $x:\mathbf{R}\rightarrow \mathbf{B}^{n}$ at
the interval $I\subset \mathbf{R}$ is denoted by $x_{|I}.$
\end{notation}

\begin{definition}
The \textbf{initial value} $x(-\infty +0)\in \mathbf{B}^{n}$ and the \textbf{%
final value} $x(\infty -0)\in \mathbf{B}^{n}$ of $x:\mathbf{R}\rightarrow 
\mathbf{B}^{n}$ are defined by%
\begin{equation*}
\exists t_{0}\in \mathbf{R},\forall t<t_{0},x(t)=x(-\infty +0),
\end{equation*}%
\begin{equation*}
\exists t_{f}\in \mathbf{R},\forall t>t_{f},x(t)=x(\infty -0).
\end{equation*}
\end{definition}

\begin{notation}
Denote by $\chi _{A}:\mathbf{R}\rightarrow \mathbf{B}$ the characteristic
function of the set $A\subset \mathbf{R}.$
\end{notation}

\begin{definition}
\label{Def1.5}The function $x:\mathbf{R}\rightarrow \mathbf{B}^{n}$ is
called \textbf{signal} if $(t_{k})\in Seq$ exists such that $\forall t\in 
\mathbf{R},$%
\begin{equation*}
x(t)=x(-\infty +0)\cdot \chi _{(-\infty ,t_{0})}(t)\oplus x(t_{0})\cdot \chi
_{\lbrack t_{0},t_{1})}(t)\oplus ...\oplus x(t_{k})\cdot \chi _{\lbrack
t_{k},t_{k+1})}(t)\oplus ...
\end{equation*}%
We have used the same symbols $\cdot ,\oplus $ for the laws that are induced
by those of $\mathbf{B}$. The set of the signals is denoted by $S^{(n)}$ and
instead of $S^{(1)}$ we usually write $S$.
\end{definition}

\begin{notation}
$P^{\ast }(S^{(n)})=\{X|X\subset S^{(n)},X\neq \emptyset \}.$
\end{notation}

\begin{definition}
The \textbf{left limit} $x(t-0)$ and the \textbf{left derivative} $Dx(t)$ of 
$x\in S^{(n)}$ expressed like in Definition \ref{Def1.5} are defined by%
\begin{equation*}
x(t-0)=x(-\infty +0)\cdot \chi _{(-\infty ,t_{0}]}(t)\oplus x(t_{0})\cdot
\chi _{(t_{0},t_{1}]}(t)\oplus ...\oplus x(t_{k})\cdot \chi
_{(t_{k},t_{k+1}]}(t)\oplus ...
\end{equation*}%
\begin{equation*}
Dx(t)=x(t-0)\oplus x(t).
\end{equation*}
\end{definition}

\begin{definition}
A multi-valued function $f:U\rightarrow P^{\ast }(S^{(n)}),U\in P^{\ast
}(S^{(m)})$ is called (\textbf{asynchronous}) \textbf{system}. Any $u\in U$
is called (\textbf{admissible}) \textbf{input} and the functions $x\in f(u)$
are called (\textbf{possible}) \textbf{states}.
\end{definition}

\begin{definition}
If $\forall u\in U,f(u)$ has exactly one element, then $f$ is called \textbf{%
deterministic} and we use the notation $f:U\rightarrow S^{(n)}$ of the
uni-valued functions.
\end{definition}

\begin{definition}
If $g:V\rightarrow P^{\ast }(S^{(n)}),V\in P^{\ast }(S^{(m)})$ is a system,
then any system $f:U\rightarrow P^{\ast }(S^{(n)}),U\in P^{\ast }(S^{(m)})$
with%
\begin{equation*}
U\subset V\text{ and }\forall u\in U,f(u)\subset g(u)
\end{equation*}%
is called a \textbf{subsystem} of $g$ and the notation is $f\subset g.$
\end{definition}

\begin{remark}
The concept of system originates in the modeling of the asynchronous
circuits. The multi-valued character of the cause-effect association is due
to statistical fluctuations in the fabrication process, the variations in
the ambiental temperature, the power supply etc.

Sometimes the systems are given by equations and/or inequalities. In this
case, determinism means that their solution is unique.

If the system $g$ models a circuit, then the system $f\subset g$ models the
same circuit more precisely, by restricting the set of the inputs perhaps.
\end{remark}

\section{The first concept of non-anticipation}

\begin{definition}
\label{Def61}The system $f:U\rightarrow P^{\ast }(S^{(n)}),$ $U\in P^{\ast
}(S^{(m)})$ is \textbf{non-anticipatory} if for all $u\in U$ and all $x\in
f(u)$ it satisfies one of the following statements:

a) $x$ is constant;

b) $u,x$ are both variable and we have%
\begin{equation}
\min \{t|u(t-0)\neq u(t)\}\leq \min \{t|x(t-0)\neq x(t)\},  \label{ineq1}
\end{equation}%
i.e. the first input switch is prior to the first output switch.
\end{definition}

\begin{remark}
The non-anticipation means that the system $f$ is in equilibrium, as
represented by the existence of the time interval $(-\infty ,t_{0}),$ where $%
u$ and $x$ are constant: $u_{|(-\infty ,t_{0})}=u(t_{0}-0)$ and $%
x_{|(-\infty ,t_{0})}=x(t_{0}-0);$ then the only possibility to get out of
this situation is the switch of the input.

Moisil presumes implicitly in his works \cite{bib102}, \cite{bib103} that
the models are non-anticipatory in the sense of Definition \ref{Def61}.
However his 'equilibrium', called 'rest position', is defined in the
presence of a 'network function' that does not exist here.
\end{remark}

\begin{example}
Any Boolean function $F:\mathbf{B}^{m}\rightarrow \mathbf{B}^{n}$ defines
for $d\geq 0$ a system $F_{d}:S^{(m)}\rightarrow S^{(n)},$ called 'ideal
combinational':%
\begin{equation*}
\forall u\in S^{(m)},\forall t\in \mathbf{R},F_{d}(u)(t)=F(u(t-d))
\end{equation*}%
which is non-anticipatory. First, $\forall u\in S^{(m)}$ we have $%
F_{d}(u)\in S^{(n)}$ indeed$.$ Second, for any $d,u,t_{0},$ $u_{|(-\infty
,t_{0})}=u(t_{0}-0)$ implies $F_{d}(u)_{|(-\infty ,t_{0})}=F_{d}(u)(t_{0}-0)$
(we have in this situation $u(t_{0}-0)=u(-\infty +0)$ and $%
F_{d}(u)(t_{0}-0)=F_{d}(u)(-\infty +0)$) and if $u(t_{0}-0)\neq u(t_{0}),$
then $F(u(t_{0}-d-0)),$ $F(u(t_{0}-d))$ represent two values that may be
equal or different. We infer that $x=F_{d}(u)$ fulfills one of a), b) from
Definition \ref{Def61}.
\end{example}

\begin{example}
The system $f:S\rightarrow S$,%
\begin{equation*}
\forall u\in S,f(u)=\chi _{\lbrack 0,\infty )}
\end{equation*}%
is anticipatory.
\end{example}

\begin{lemma}
\label{Lem1}If $g:V\rightarrow P^{\ast }(S^{(n)}),V\in P^{\ast }(S^{(m)})$
is a non-anticipatory system, then any system $f:U\rightarrow P^{\ast
}(S^{(n)}),U\in P^{\ast }(S^{(m)})$ with $f\subset g$ is non-anticipatory.
\end{lemma}

\begin{proof}
Let be $u\in U.$ We have the following possibilities:

i) $u$ is constant. From Definition \ref{Def61} we have that $\forall x\in
g(u),x$ is constant, in particular $\forall x\in f(u),x$ is constant.
Therefore, $f$ is non-anticipatory;

ii) $u$ is variable. Let $x\in f(u)$ be arbitrary. Then

ii.1) $x$ is constant implies that $f$ is non-anticipatory, by Definition %
\ref{Def61} a);

ii.2) $x$ is variable. As element of $g(u),x$ satisfies (\ref{ineq1}) and,
by Definition \ref{Def61} b), $f$ is non-anticipatory.
\end{proof}

\section{Choosing 0 as initial time instant}

\begin{notation}
$\forall d\in \mathbf{R},$ the function $\tau ^{d}:\mathbf{R}\rightarrow 
\mathbf{R}$ is the translation with $d$, thus for any $x\in S^{(n)}$ we
denote by $x\circ \tau ^{d}$ the function%
\begin{equation*}
\forall t\in \mathbf{R},(x\circ \tau ^{d})(t)=x(t-d).
\end{equation*}
\end{notation}

\begin{definition}
The system $f$ is \textbf{time invariant} if $\forall d\in \mathbf{R}%
,\forall u\in U,$%
\begin{equation*}
u\circ \tau ^{d}\in U,
\end{equation*}%
\begin{equation*}
\forall x\in f(u),x\circ \tau ^{d}\in f(u\circ \tau ^{d}).
\end{equation*}
\end{definition}

\begin{notation}
\label{Not21}We use the notation%
\begin{equation*}
S_{0}^{(m)}=\{u|u\in S^{(m)},\forall t<0,u(t)=u(-\infty +0)\}.
\end{equation*}
\end{notation}

\begin{theorem}
\label{The4.4}We state the following properties relative to some system $%
\widehat{f}:\widehat{U}\rightarrow P^{\ast }(S^{(n)}),$ $\widehat{U}\in
P^{\ast }(S^{(m)}):$

i) $\widehat{U}\subset S_{0}^{(m)};$

ii) $\forall u\in \widehat{U},\widehat{f}(u)\subset S_{0}^{(n)};$

iii) $\forall d\in \mathbf{R},\forall u\in \widehat{U},\forall x,$%
\begin{equation*}
(x\in \widehat{f}(u)\;and\;u\circ \tau ^{d}\in \widehat{U})\Longrightarrow
x\circ \tau ^{d}\in \widehat{f}(u\circ \tau ^{d}).
\end{equation*}

a) The time-invariant non-anticipatory system $f:U\rightarrow P^{\ast
}(S^{(n)}),U\in P^{\ast }(S^{(m)})$ is given. We define the system $\widehat{%
f}:\widehat{U}\rightarrow P^{\ast }(S^{(n)})$ by%
\begin{equation*}
\widehat{U}=\{u|u\in U\cap S_{0}^{(m)}\;and\;f(u)\cap S_{0}^{(n)}\neq
\emptyset \},
\end{equation*}%
\begin{equation}
\forall u\in \widehat{U},\widehat{f}(u)=f(u)\cap S_{0}^{(n)}.  \label{Cho1}
\end{equation}%
The system $\widehat{f}$ fulfills i), ii), iii) and is also non-anticipatory.

b) Let be the system $\widehat{f}:\widehat{U}\rightarrow P^{\ast }(S^{(n)})$
satisfying the properties i), ii), iii) and non-anticipation. The system $%
f:U\rightarrow P^{\ast }(S^{(n)}),U\in P^{\ast }(S^{(m)})$ defined by%
\begin{equation*}
U=\{u\circ \tau ^{d}|d\in \mathbf{R},u\in \widehat{U}\},
\end{equation*}%
\begin{equation*}
\forall d\in \mathbf{R},\forall u\in \widehat{U},f(u\circ \tau
^{d})=\{x\circ \tau ^{d}|x\in \widehat{f}(u)\}
\end{equation*}%
is time invariant and non-anticipatory.
\end{theorem}

\begin{proof}
a) We show that $U\cap S_{0}^{(m)}\neq \emptyset .$ Let be $u\in U.$ We have
the possibilities:

1) $u$ is constant. Then $u\in S_{0}^{(m)}$, thus $u\in U\cap S_{0}^{(m)};$

2) $u$ is variable.

We denote $d=\min \{t|u(t-0)\neq u(t)\}.$ If $d\geq 0$, then $u\in
S_{0}^{(m)}$ and $u\in U\cap S_{0}^{(m)}$ are true. If $d<0$, then for any $%
d^{\prime }\geq -d$, $u\circ \tau ^{d^{\prime }}\in U$ is true because $U$
is invariant to translations and $u\circ \tau ^{d^{\prime }}\in S_{0}^{(m)}$
holds true also, making $u\circ \tau ^{d^{\prime }}\in U\cap S_{0}^{(m)}$
true.

We show that $\widehat{U}\neq \emptyset .$ We take some arbitrary $u\in
U\cap S_{0}^{(m)}$. If $f(u)\cap S_{0}^{(n)}\neq \emptyset ,$ the property
is true, otherwise let be some $x\in f(u)$. The fact that $x\notin
S_{0}^{(n)}$ shows that it is variable and if we denote by $d=\min
\{t|x(t-0)\neq x(t)\},$ we have $d<0$. Remark that for all $d^{\prime }\geq
-d,$ $u\circ \tau ^{d^{\prime }}\in U,$ $u\circ \tau ^{d^{\prime }}\in
S_{0}^{(m)},$ $x\circ \tau ^{d^{\prime }}\in f(u\circ \tau ^{d^{\prime }})$
and $x\circ \tau ^{d^{\prime }}\in S_{0}^{(n)}$ take place. In other words $%
u\circ \tau ^{d^{\prime }}\in \widehat{U}.$

This shows that $\widehat{f}$ is well defined, in the sense that $\widehat{U}%
\neq \emptyset $ and $\forall u\in \widehat{U},\widehat{f}(u)\neq \emptyset $%
. Moreover, i) and ii) are obviously satisfied.

We show now the truth of iii). We take $d\in \mathbf{R},$ $u\in \widehat{U},$
$x$ arbitrary with $x\in \widehat{f}(u)$ and $u\circ \tau ^{d}\in \widehat{U}
$ true. We have the possibilities:

j) $x$ is constant. Then $x\circ \tau ^{d}=x$ is constant and $x\circ \tau
^{d}\in S_{0}^{(n)}$;

jj) $x$ is variable. Because $x\in f(u),$ from the non-anticipation of $f$
we have that $u$ is variable and%
\begin{equation*}
0\leq \min \{t|(u\circ \tau ^{d})(t-0)\neq (u\circ \tau ^{d})(t)\}\leq \min
\{t|(x\circ \tau ^{d})(t-0)\neq (x\circ \tau ^{d})(t)\},
\end{equation*}%
showing that $x\circ \tau ^{d}\in S_{0}^{(n)}$.

In both cases j), jj), $x\in \widehat{f}(u)$ has implied $x\in f(u)$ and,
furthermore, $x\circ \tau ^{d}\in f(u\circ \tau ^{d})$ from the time
invariance of $f$ and, eventually, $x\circ \tau ^{d}\in \widehat{f}(u\circ
\tau ^{d})$ $(=f(u\circ \tau ^{d})\cap S_{0}^{(n)})$.

Because $\widehat{f}\subset f,$\ the non-anticipation of $\widehat{f}$\ is a
consequence of Lemma \ref{Lem1}.

b) We show that $f$ is well defined in the sense that if $d,d^{\prime }\in 
\mathbf{R}$ and $u,v\in \widehat{U}$ satisfy $u\circ \tau ^{d}=v\circ \tau
^{d^{\prime }},$ we get $f(u\circ \tau ^{d})=f(v\circ \tau ^{d^{\prime }}).$
Let be $x\circ \tau ^{d}\in f(u\circ \tau ^{d}).$ We infer that $x\in 
\widehat{f}(u)$ and $v=u\circ \tau ^{d-d^{\prime }}\in \widehat{U}.$ From
iii) we have that $x\circ \tau ^{d-d^{\prime }}\in \widehat{f}(v),$ i.e. $%
x\circ \tau ^{d}=x\circ \tau ^{d-d^{\prime }}\circ \tau ^{d^{\prime }}\in
f(v\circ \tau ^{d^{\prime }}).$ We have obtained that $f(u\circ \tau
^{d})\subset f(v\circ \tau ^{d^{\prime }})$ and the inverse inclusion is
shown similarly.

We show that $U$ is invariant to translations. Let be $v\in U$. Then there
are some $d\in \mathbf{R}$ and $u\in \widehat{U}$ such that $v=u\circ \tau
^{d}$. For an arbitrary $d^{\prime }\in \mathbf{R}$, as $v\circ \tau
^{d^{\prime }}=u\circ \tau ^{d+d^{\prime }}$, we infer $v\circ \tau
^{d^{\prime }}\in U$.

We show that $f$ is time invariant. Let be $v\in U$ and $y\in f(v)$, meaning
the existence of $u\in \widehat{U}$ and $d\in \mathbf{R}$ with $v=u\circ
\tau ^{d}$. We get $y\in f(u\circ \tau ^{d})=\{x\circ \tau ^{d}|x\in 
\widehat{f}(u)\}$. In other words $\exists x,y=x\circ \tau ^{d}$ and $x\in 
\widehat{f}(u)$. We take an arbitrary $d^{\prime }\in \mathbf{R}$ for which $%
y\circ \tau ^{d^{\prime }}=x\circ \tau ^{d+d^{\prime }}$, $y\circ \tau
^{d^{\prime }}\in \{x\circ \tau ^{d+d^{\prime }}|x\in \widehat{f}%
(u)\}=f(u\circ \tau ^{d+d^{\prime }})=f(v\circ \tau ^{d^{\prime }})$.

We show now that $f$ is non-anticipatory. Let us take, like previously, $%
v\in U$ and $y\in f(v)$, for which there are $u\in \widehat{U},x\in \widehat{%
f}(u)$ and $d\in \mathbf{R}$ such that $v=u\circ \tau ^{d}$ and $y=x\circ
\tau ^{d}$. We have the possibilities:

I) $y$ is constant. Then $f$ is non-anticipatory;

II) $y$ is variable. Then $x\in \widehat{f}(u)$ is variable and the
hypothesis concerning the non-anticipation of $\widehat{f}$ states that $u$
is variable and%
\begin{equation*}
\min \{t|u(t-0)\neq u(t)\}\leq \min \{t|x(t-0)\neq x(t)\}.
\end{equation*}%
We obtain%
\begin{equation*}
\min \{t|v(t-0)\neq v(t)\}=\min \{t|(u\circ \tau ^{d})(t-0)\neq (u\circ \tau
^{d})(t)\}=
\end{equation*}%
\begin{equation*}
=d+\min \{t|u(t-0)\neq u(t)\}\leq d+\min \{t|x(t-0)\neq x(t)\}=
\end{equation*}%
\begin{equation*}
=\min \{t|(x\circ \tau ^{d})(t-0)\neq (x\circ \tau ^{d})(t)\}=\min
\{t|y(t-0)\neq y(t)\}.
\end{equation*}
\end{proof}

\begin{remark}
$S_{0}^{(m)}$ consists in these signals $u\in S^{(m)}$ that accept the
'initial time instant' $t_{0}$ be 0. Items i), ii) of Theorem \ref{The4.4}
mean that the inputs and the states of $\widehat{f}$ accept the initial time
instant be $0$ and item iii) of that theorem represents time invariance
adapted to the situation when $\widehat{U}$ is not closed to translations
(any $\widehat{U}\subset S_{0}^{(m)}$ that contains non-constant signals is
not invariant to translations).

The possibility of choosing $0$ as initial time instant simplifies a little
the study of the asynchronous systems.
\end{remark}

\section{Non-Anticipation, the Second Definition}

\begin{definition}
\label{Def62}Let the system $f:U\rightarrow P^{\ast }(S^{(n)})$ be given$,$ $%
U\in P^{\ast }(S^{(m)})$. It is called \textbf{non-anticipatory} if $\forall
t\in \mathbf{R},$ $\forall u\in U,$ $\forall v\in U,$%
\begin{equation*}
u_{|(-\infty ,t)}=v_{|(-\infty ,t)}\Longrightarrow \{x_{|(-\infty ,t]}|x\in
f(u)\}=\{y_{|(-\infty ,t]}|y\in f(v)\}.
\end{equation*}
\end{definition}

\begin{remark}
Definition \ref{Def62} states that the history of all the possible states
until the present moment, including the present depends only on the history
of the input and it does not depend on the present and the future values of
the input. The definition means that $\forall t\in \mathbf{R}$ a function $%
f_{t}$ exists that associates $\forall u\in U$ to $u_{|(-\infty ,t)}$ the set%
\begin{equation*}
f_{t}(u_{|(-\infty ,t)})=\{x_{|(-\infty ,t]}|x\in f(u)\}.
\end{equation*}

Definition \ref{Def62} represents a perspective of non-anticipation, other
than the previous one and the two properties are logically independent.
\end{remark}

\begin{example}
The deterministic system $f:S^{(m)}\rightarrow S,$%
\begin{equation*}
\forall u\in S^{(m)},f(u)=\chi _{\lbrack 0,1)}\oplus (u_{1}\circ \tau
^{1})\cdot \chi _{\lbrack 1,\infty )}
\end{equation*}%
is non-anticipatory in the sense of Definition \ref{Def62}. The system $f$
is anticipatory in the sense of Definition \ref{Def61} because for $%
u_{1}=\chi _{\lbrack 2,\infty )}$, $u_{2}=...=u_{m}=0$ the contradiction $%
\min \{t|u(t-0)\neq u(t)\}=2>0=\min \{t|x(t-0)\neq x(t)\}$ is obtained.
\end{example}

\begin{example}
The deterministic system $f:S\rightarrow S,$%
\begin{equation*}
\forall u\in S,f(u)=\left\{ 
\begin{array}{c}
1,if\;u=\chi _{\lbrack 0,\infty )} \\ 
u,\;otherwise\quad%
\end{array}%
\right. ,
\end{equation*}%
is anticipatory in the sense of Definition \ref{Def62} because for $t=1,$ $%
u=\chi _{\lbrack 0,\infty )},$ $v=\chi _{\lbrack 0,2)}$ we have $%
u_{|(-\infty ,1)}=v_{|(-\infty ,1)}$ but $1_{|(-\infty ,1]}\neq \chi
_{\lbrack 0,2)_{|(-\infty ,1]}}$. However it is non-anticipatory in the
sense of Definition \ref{Def61}.
\end{example}

\begin{example}
The deterministic system $f:S\rightarrow S$%
\begin{equation*}
\forall u\in S,f(u)=\left\{ 
\begin{array}{c}
1,if\;u=\chi _{\lbrack 0,\infty )}\quad \quad \\ 
u\circ \tau ^{-1},\;otherwise%
\end{array}%
\right.
\end{equation*}%
is anticipatory in the sense of both Definitions \ref{Def61} and \ref{Def62}.
\end{example}

\begin{example}
\label{Exa62}The deterministic system%
\begin{equation*}
Dx(t)=(x(t-0)\oplus u(t-0))\cdot \overline{\underset{\xi \in (t-d,t)}{%
\bigcup }Du(\xi )}
\end{equation*}%
$u,x\in S,d>0$ is non-anticipatory in the sense of both Definitions \ref%
{Def61}, \ref{Def62}. The idea expressed by such an equation is: $x$
switches ($Dx(t)=1$) at these time instants when $u$ has indicated the
necessity of such a switch $(x(t-0)\oplus u(t-0)=1)$ for $d$ time units ($%
u_{|[t-d,t)}$ is the constant function, with null derivative in the interval 
$(t-d,t)$). This equation models the delay circuit.
\end{example}

\section{Other Definitions of Non-Anticipation. Non-Anticipation$^{\ast }$}

\begin{definition}
\label{Def63}Let be the system $f:U\rightarrow P^{\ast }(S^{(n)}),$ $U\in
P^{\ast }(S^{(m)}).$ It is called \textbf{non-anticipatory} if it satisfies
one of the following conditions, called conditions of non-anticipation:

i) $\forall t\in \mathbf{R},\forall u\in U,\forall v\in U,$%
\begin{equation*}
u_{|(-\infty ,t)}=v_{|(-\infty ,t)}\Longrightarrow \{x(t)|x\in
f(u)\}=\{y(t)|y\in f(v)\};
\end{equation*}

ii) $\forall t\in \mathbf{R},\forall u\in U,\forall v\in U,\exists d>0,$%
\begin{equation*}
u_{|[t-d,t)}=v_{|[t-d,t)}\Longrightarrow \{x(t)|x\in f(u)\}=\{y(t)|y\in
f(v)\};
\end{equation*}

iii) $\forall t\in \mathbf{R},\exists d>0,\forall u\in U,\forall v\in U,$%
\begin{equation*}
u_{|[t-d,t)}=v_{|[t-d,t)}\Longrightarrow \{x(t)|x\in f(u)\}=\{y(t)|y\in
f(v)\};
\end{equation*}

iv) $\exists d>0,\forall t\in \mathbf{R},\forall u\in U,\forall v\in U,$%
\begin{equation*}
u_{|[t-d,t)}=v_{|[t-d,t)}\Longrightarrow \{x(t)|x\in f(u)\}=\{y(t)|y\in
f(v)\};
\end{equation*}

v) $\forall t\in \mathbf{R},\forall u\in U,\forall v\in U,$%
\begin{equation*}
u_{|(-\infty ,t]}=v_{|(-\infty ,t]}\Longrightarrow \{x_{|(-\infty ,t]}|x\in
f(u)\}=\{y_{|(-\infty ,t]}|y\in f(v)\};
\end{equation*}

vi) $\forall t\in \mathbf{R},\forall u\in U,\forall v\in U,$%
\begin{equation*}
u_{|(-\infty ,t]}=v_{|(-\infty ,t]}\Longrightarrow \{x(t)|x\in
f(u)\}=\{y(t)|y\in f(v)\};
\end{equation*}

vii) $\forall t\in \mathbf{R},\forall u\in U,\forall v\in U$,$\exists
d,\exists d^{\prime },0\leq d\leq d^{\prime }$ and%
\begin{equation*}
u_{|[t-d^{\prime },t-d]}=v_{|[t-d^{\prime },t-d]}\Longrightarrow \{x(t)|x\in
f(u)\}=\{y(t)|y\in f(v)\};
\end{equation*}

viii) $\forall t\in \mathbf{R},\exists d,\exists d^{\prime },0\leq d\leq
d^{\prime }$ and $\forall u\in U,\forall v\in U$,%
\begin{equation*}
u_{|[t-d^{\prime },t-d]}=v_{|[t-d^{\prime },t-d]}\Longrightarrow \{x(t)|x\in
f(u)\}=\{y(t)|y\in f(v)\};
\end{equation*}

ix) $\exists d,\exists d^{\prime },0\leq d\leq d^{\prime }$ and $\forall
t\in \mathbf{R},\forall u\in U,\forall v\in U,$%
\begin{equation*}
u_{|[t-d^{\prime },t-d]}=v_{|[t-d^{\prime },t-d]}\Longrightarrow \{x(t)|x\in
f(u)\}=\{y(t)|y\in f(v)\}.
\end{equation*}
\end{definition}

\begin{theorem}
\label{The194}If $f:U\rightarrow S^{(n)}$ is a deterministic system, then
Definition \ref{Def63} v) and Definition \ref{Def63} vi) are equivalent. We
have that Definition \ref{Def62} and Definition \ref{Def63} i) are
equivalent in this case too.
\end{theorem}

\begin{proof}
We prove the first statement. Because v)$\Longrightarrow $vi) is obvious, we
prove vi)$\Longrightarrow $v). Let us suppose against all reason that v) is
not true, i.e. $\exists t\in \mathbf{R},\exists u\in U,\exists v\in
U,u_{|(-\infty ,t]}=v_{|(-\infty ,t]}$ and $f(u)_{|(-\infty ,t]}\neq
f(v)_{|(-\infty ,t]}$. This means the existence of $t^{\prime }\leq t$ such
that $u_{|(-\infty ,t^{\prime }]}=v_{|(-\infty ,t^{\prime }]}$ and $%
f(u)(t^{\prime })\neq f(v)(t^{\prime }),$ contradiction with vi).
\end{proof}

\begin{remark}
\label{Rem51}In Definition \ref{Def63}, all of i),...,ix) express the same
idea like Definition \ref{Def62}, namely that the present depends on the
past only and it is independent on the future. The implications are:%
\begin{equation*}
\begin{array}{ccccccccc}
iv) & \Longrightarrow  & iii) & \Longrightarrow  & ii) & \Longrightarrow  & 
i) & \Longleftarrow  & Definition\text{ \ref{Def62}} \\ 
&  &  &  &  &  & \Downarrow  &  & \Downarrow  \\ 
ix) & \Longrightarrow  & viii) & \Longrightarrow  & vii) & \Longrightarrow 
& vi) & \Longleftarrow  & v)%
\end{array}%
\end{equation*}

In ii),...,iv), vii),...,ix) the boundness of the memory occurs: these are
systems whose states do not depend on all the input segment $u_{|(-\infty
,t)},$ but on the last $d$ time units $u_{|[t-d,t)}$ only and similarly for $%
u_{|(-\infty ,t]}$ and $u_{|[t-d^{\prime },t-d]}.$

Now have a look at the non-anticipation property iv). We note that if $d>0$
is a number for which it is fulfilled, then any number $d^{\prime }\geq d$
fulfills it also: $\forall t\in \mathbf{R},\forall u\in U,\forall v\in U,$%
\begin{equation*}
u_{|[t-d^{\prime },t)}=v_{|[t-d^{\prime },t)}\Longrightarrow \{x(t)|x\in
f(u)\}=\{y(t)|y\in f(v)\}.
\end{equation*}%
Our problem is whether, for a system $f$, the set of those $d$ satisfying
implication iv) is bounded from below by some $d^{\prime \prime }>0,$
because we have a non-anticipation property

$\forall t\in \mathbf{R},\forall u\in U,\forall v\in U,$%
\begin{equation*}
u(t-0)=v(t-0)\Longrightarrow \{x(t)|x\in f(u)\}=\{y(t)|y\in f(v)\}
\end{equation*}%
also, like in the example%
\begin{equation*}
u(t-0)\cdot x(t)=0
\end{equation*}%
where $u,x\in S$. If this lower bound exists, we obtain a new shading of
that concept of non-anticipation. The problem of the existence of such
bounds is, in principle, the same if $d$ is variable like in ii), iii) or if
instead of one parameter $d$ we have two parameters $d,d^{\prime }$ and two
bounds, like in vii), viii), ix).

Remark that the reasoning of Theorem \ref{The194} is impossible to use if $f$
is non-deterministic. We suppose, for this, that the system $f:S\rightarrow
P^{\ast }(S)$ satisfies $f(0)=\{0,1\},$ $f(\chi _{\lbrack 2,\infty
)})=\{\chi _{(-\infty ,0)},\chi _{\lbrack 0,\infty )}\},$ where $0,1\in S$
are the constant functions. We have $\forall t\in \lbrack 0,2),$%
\begin{equation*}
0_{|(-\infty ,t]}=\chi _{\lbrack 2,\infty )}{}_{|(-\infty ,t]}\text{ and }
\end{equation*}%
\begin{equation*}
\text{and }\{0_{|(-\infty ,t]},1_{|(-\infty ,t]}\}\neq \{\chi _{(-\infty
,0)}{}_{|(-\infty ,t]},\chi _{\lbrack 0,\infty )}{}_{|(-\infty ,t]}\}\text{
and}
\end{equation*}%
\begin{equation*}
\text{and }\{x(t)|x\in f(0)\}=\{0,1\}=\{y(t)|y\in f(\chi _{\lbrack 2,\infty
)})\}.
\end{equation*}
\end{remark}

\begin{example}
The system $I_{d}:S\rightarrow S$ called the 'pure delay model' of the delay
circuit, defined by $\forall u\in S,x(t)=I_{d}(u)(t)=u(t-d),$ satisfies for $%
d>0$ all the non-anticipation properties i),...,ix) from Definition \ref%
{Def63}.
\end{example}

\begin{example}
\label{Exa63}Let the system $f:S\rightarrow P^{\ast }(S)$ (version of the
'bounded delay model' of the delay circuit) be defined by the inequalities%
\begin{equation*}
\underset{\xi \in \lbrack t-d_{r},t)}{\bigcap }u(\xi )\leq x(t)\leq \underset%
{\xi \in \lbrack t-d_{f},t)}{\bigcup }u(\xi ),
\end{equation*}%
where $d_{r}>0,d_{f}>0.$ It satisfies all the non-anticipation properties
i),...,ix) from Definition \ref{Def63}.
\end{example}

\begin{example}
The system $f:S\rightarrow P^{\ast }(S)$ (version of the 'bounded delay
model' of the delay circuit) described by the inequalities%
\begin{equation*}
\underset{\xi \in \lbrack t-d^{\prime },t-d]}{\bigcap }u(\xi )\leq x(t)\leq 
\underset{\xi \in \lbrack t-d^{\prime },t-d]}{\bigcup }u(\xi ),
\end{equation*}%
where $0\leq d\leq d^{\prime }$ satisfies the non-anticipation properties
v),...,ix) from Definition \ref{Def63}.
\end{example}

\begin{example}
The system $f:S\rightarrow P^{\ast }(S)$ defined by%
\begin{equation*}
\int\limits_{-\infty }^{t}Du\leq x(t)
\end{equation*}%
where%
\begin{equation*}
\int\limits_{-\infty }^{t}Du=\left\{ 
\begin{array}{c}
1,\;if\;|suppDu\cap (-\infty ,t]|\;is\;odd\; \\ 
0,\;if\;|suppDu\cap (-\infty ,t]|\;is\;even%
\end{array}%
\right.
\end{equation*}%
satisfies the non-anticipation properties v), vi) of Definition \ref{Def63}.
We have denoted by $|suppDu\cap (-\infty ,t]|$ the number of elements of the
finite set $\{\xi |\xi \in \mathbf{R},Du(\xi )=1\}\cap (-\infty ,t]$ and we
have supposed that $0$ is an even number.
\end{example}

\begin{example}
Denote by $\varphi :S^{(m)}\rightarrow \lbrack 0,\infty )$ the function $%
\forall u\in S^{(m)},$%
\begin{equation*}
\varphi (u)=\left\{ 
\begin{array}{c}
0,\text{if}\;u\;\text{is\ constant\quad \quad \quad \quad \quad \quad \quad
\quad \quad \quad } \\ 
\max \{-\min \{t|u(t-0)\neq u(t)\},\min \{t|u(t-0)\neq u(t)\}\},\text{if}%
\;u\;\text{is\ variable}%
\end{array}%
\right. 
\end{equation*}%
The deterministic system%
\begin{equation*}
x(t)=\underset{\xi \in \lbrack t-2\varphi (u),t-\varphi (u)]}{\bigcap }u(\xi
),
\end{equation*}%
$u,x\in S,$ satisfies the non-anticipation property vii) of Definition \ref%
{Def63}.
\end{example}

\begin{definition}
\label{Def64}The system $f$ is called \textbf{non-anticipatory}$^{\ast }$ if
it satisfies one of the following conditions, called conditions of
non-anticipation$^{\ast }:$

i) $\forall t\in \mathbf{R},\forall u\in U,\forall v\in U,$%
\begin{equation*}
(u_{|[t,\infty )}=v_{|[t,\infty )}\text{ and }\{x(t)|x\in f(u)\}=\{y(t)|y\in
f(v)\})\Longrightarrow
\end{equation*}%
\begin{equation*}
\Longrightarrow \{x_{|[t,\infty )}|x\in f(u)\}=\{y_{|[t,\infty )}|y\in
f(v)\};
\end{equation*}

ii) $\forall t\in \mathbf{R},\forall u\in U,\forall v\in U,$%
\begin{equation*}
u_{|[t,\infty )}=v_{|[t,\infty )}\Longrightarrow \exists t^{\prime }\in 
\mathbf{R},\{x_{|[t^{\prime },\infty )}|x\in f(u)\}=\{y_{|[t^{\prime
},\infty )}|y\in f(v)\};
\end{equation*}

iii) $\forall t\in \mathbf{R},\forall u\in U,\forall v\in U,$%
\begin{equation*}
(u_{|[t,\infty )}=v_{|[t,\infty )}\text{ and }\{x_{|(-\infty ,t]}|x\in
f(u)\}=\{y_{|(-\infty ,t]}|y\in f(v)\})\Longrightarrow
\end{equation*}%
\begin{equation*}
\Longrightarrow \exists t^{\prime }\in \mathbf{R},\{x_{|[t^{\prime },\infty
)}|x\in f(u)\}=\{y_{|[t^{\prime },\infty )}|y\in f(v)\}.
\end{equation*}
\end{definition}

\begin{remark}
We remark that property i) resembles somehow with fixing the initial
conditions in a differential equation ($\{x(t)|x\in f(u)\}=\{y(t)|y\in
f(v)\} $). The consequence is that the solution is unique ($\{x_{|[t,\infty
)}|x\in f(u)\}=\{y_{|[t,\infty )}|y\in f(v)\}$) under an arbitrary given
input ($u_{|[t,\infty )}=v_{|[t,\infty )}$).

The reader is invited to write other similar properties of non-anticipation
and non-anticipation$^{\ast }.$
\end{remark}

\section{The Transfers of the Non-Anticipatory Systems}

\begin{theorem}
\label{The207}Let the system $f$ satisfy the conditions:

a) $U$ is closed under 'concatenation' $\forall t\in \mathbf{R},\forall u\in
U,\forall v\in U,$%
\begin{equation*}
u\cdot \chi _{(-\infty ,t)}\oplus v\cdot \chi _{\lbrack t,\infty )}\in U;
\end{equation*}

b) non-anticipation $\forall t\in \mathbf{R},$ $\forall u\in U,$ $\forall
v\in U,$%
\begin{equation*}
u_{|(-\infty ,t)}=v_{|(-\infty ,t)}\Longrightarrow \{x_{|(-\infty ,t]}|x\in
f(u)\}=\{y_{|(-\infty ,t]}|y\in f(v)\};
\end{equation*}

c) non-anticipation$^{\ast }$ $\forall t\in \mathbf{R},$ $\forall u\in U,$ $%
\forall v\in U,$%
\begin{equation*}
(u_{|[t,\infty )}=v_{|[t,\infty )}\text{ and }\{x(t)|x\in f(u)\}=\{y(t)|y\in
f(v)\})\Longrightarrow
\end{equation*}%
\begin{equation*}
\Longrightarrow \{x_{|[t,\infty )}|x\in f(u)\}=\{y_{|[t,\infty )}|y\in
f(v)\};
\end{equation*}

d) time invariance $\forall d\in \mathbf{R},\forall u\in U,$%
\begin{equation*}
u\circ \tau ^{d}\in U,
\end{equation*}%
\begin{equation*}
\forall x\in f(u),x\circ \tau ^{d}\in f(u\circ \tau ^{d});
\end{equation*}

e) $t_{1},$ $t_{2}\in \mathbf{R},$ $u^{0},u^{1}\in U$ and $\mu ,\mu ^{\prime
},\mu ^{\prime \prime }\in \mathbf{B}^{n}$ are given such that%
\begin{equation}
\forall x\in f(u^{0}),\exists t_{0}<t_{1},x(t_{0})=\mu ,  \label{tna1}
\end{equation}%
\begin{equation}
\forall x\in f(u^{0}),x(t_{1})=\mu ^{\prime },  \label{tna1a}
\end{equation}%
\begin{equation}
\forall x^{\prime }\in f(u^{1}),x^{\prime }(t_{2})=\mu ^{\prime },
\label{tna2}
\end{equation}%
\begin{equation}
\forall x^{\prime }\in f(u^{1}),\exists t_{3}>t_{2},x^{\prime }(t_{3})=\mu
^{\prime \prime }.  \label{tna2a}
\end{equation}%
Put $d=t_{1}-t_{2}.$ Then $\widetilde{u}\in U$ defined as%
\begin{equation}
\widetilde{u}=u^{0}\cdot \chi _{(-\infty ,t_{1})}\oplus (u^{1}\circ \tau
^{d})\cdot \chi _{\lbrack t_{1},\infty )},  \label{tna4}
\end{equation}%
satisfies%
\begin{equation}
\forall \widetilde{x}\in f(\widetilde{u}),\exists t_{0}<t_{1},\widetilde{x}%
(t_{0})=\mu ,  \label{tna3}
\end{equation}%
\begin{equation}
\forall \widetilde{x}\in f(\widetilde{u}),\exists t_{3}^{\prime }>t_{1},%
\widetilde{x}(t_{3}^{\prime })=\mu ^{\prime \prime }.  \label{tna3a}
\end{equation}
\end{theorem}

\begin{proof}
$\widetilde{u}$ belongs to $U$ indeed, because of a) and d). We remark that
we have%
\begin{equation}
\widetilde{u}_{|(-\infty ,t_{1})}=u_{|(-\infty ,t_{1})}^{0}.  \label{tna5}
\end{equation}%
From (\ref{tna5}) and b) we infer%
\begin{equation}
\{\widetilde{x}_{|(-\infty ,t_{1}]}|\widetilde{x}\in f(\widetilde{u}%
)\}=\{x_{|(-\infty ,t_{1}]}|x\in f(u^{0})\}  \label{tna6}
\end{equation}%
and if, in addition, we take into account (\ref{tna1}), (\ref{tna1a}), then
we get the truth of (\ref{tna3}) and of%
\begin{equation}
\forall \widetilde{x}\in f(\widetilde{u}),\widetilde{x}(t_{1})=\mu ^{\prime
}.  \label{tna8}
\end{equation}%
Let be now some arbitrary $x^{\prime \prime }\in f(u^{1}\circ \tau ^{d}).$
From d) we obtain the existence of $x^{\prime }\in f(u^{1}),$ such that $%
x^{\prime \prime }=x^{\prime }\circ \tau ^{d}$ (namely $x^{\prime
}=x^{\prime \prime }\circ \tau ^{-d}$) and we have $x^{\prime \prime
}(t_{1})=(x^{\prime }\circ \tau ^{d})(t_{1})=x^{\prime }(t_{2})=\mu ^{\prime
}$ (we have taken into account (\ref{tna2})) thus%
\begin{equation}
\forall x^{\prime \prime }\in f(u^{1}\circ \tau ^{d}),x^{\prime \prime
}(t_{1})=\mu ^{\prime }  \label{tna9}
\end{equation}%
and, similarly,%
\begin{equation}
\forall x^{\prime \prime }\in f(u^{1}\circ \tau ^{d}),\exists t_{3}^{\prime
}>t_{1},x^{\prime \prime }(t_{3}^{\prime })=\mu ^{\prime \prime }.
\label{tna10}
\end{equation}%
We see that%
\begin{equation}
\widetilde{u}_{|[t_{1},\infty )}=(u^{1}\circ \tau ^{d})_{|[t_{1},\infty )}.
\label{tna11}
\end{equation}%
The hypothesis of c) is fulfilled by $t_{1},$ $\widetilde{u}$ and $%
u^{1}\circ \tau ^{d},$ as follows from (\ref{tna8}), (\ref{tna9}) and (\ref%
{tna11}). The conclusion of c) expresses the fact that%
\begin{equation}
\{\widetilde{x}_{|[t_{1},\infty )}|\widetilde{x}\in f(\widetilde{u}%
)\}=\{x_{|[t_{1},\infty )}^{\prime \prime }|x^{\prime \prime }\in
f(u^{1}\circ \tau ^{d})\}
\end{equation}%
and, by (\ref{tna10}), we get the truth of (\ref{tna3a}).
\end{proof}

\begin{remark}
The relations (\ref{tna1}), (\ref{tna2a}), (\ref{tna3}), (\ref{tna3a}) show
the asynchronous access (weaker, the time instant when the access happens
depends on $x$) of the states of $f$ to the values $\mu ,\mu ^{\prime \prime
}$ and the relations (\ref{tna1a}), (\ref{tna2}) represent the synchronous
access (stronger, the time instant when the access happens is the same for
all $x$; these two accesses must match) of the states of $f$ to the value $%
\mu ^{\prime }.$ The theorem states that if $f(u^{0})$ transfers $\mu $ in $%
\mu ^{\prime }$ and $f(u^{1})$ transfers $\mu ^{\prime }$ in $\mu ^{\prime
\prime },$ then $f(\widetilde{u})$ transfers $\mu $ in $\mu ^{\prime \prime
}.$

Several versions of this theorem are obtained if we take $t_{1}=t_{2}$ (then
time invariance disappears from the hypothesis), if we have in the
hypothesis countable many transfers (instead of two; these transfers must
have synchronous accesses), if we state in the hypothesis a
controllability/accessibility request etc.
\end{remark}

\section{The Fundamental Mode}

\begin{definition}
Consider the system $f$ supposed to be non-anticipatory (Definition \ref%
{Def62}) and let $u\in U$ be a fixed input. If there are $(t_{k})\in Seq,$ $%
(u^{k})\in U$ and $(\mu ^{k})\in \mathbf{B}^{n}$ such that%
\begin{equation*}
\forall x\in f(u^{0}),x_{|(-\infty ,t_{0})}=\mu ^{0}\;and\;x_{|[t_{1},\infty
)}=\mu ^{1},
\end{equation*}%
\begin{equation*}
u_{|(-\infty ,t_{1})}=u_{|(-\infty ,t_{1})}^{0},\;u_{|(-\infty
,t_{2})}=u_{|(-\infty ,t_{2})}^{1},\;u_{|(-\infty ,t_{3})}=u_{|(-\infty
,t_{3})}^{2},...
\end{equation*}%
\begin{equation*}
\forall x\in f(u^{1}),x_{|[t_{2},\infty )}=\mu ^{2},\;\forall x\in
f(u^{2}),x_{|[t_{3},\infty )}=\mu ^{3},\;\forall x\in
f(u^{3}),x_{|[t_{4},\infty )}=\mu ^{4},...
\end{equation*}%
then the input $u$ is called a \textbf{fundamental} (\textbf{operating}) 
\textbf{mode} (\textbf{of} $f$).
\end{definition}

\begin{remark}
The evolution of $f$ under the fundamental mode $u$ may be interpreted as
the discrete time evolution of a deterministic system of the form%
\begin{equation*}
\mu ^{0}=x(0)\overset{u^{0}}{\rightarrow }\mu ^{1}=x(1)\overset{u^{1}}{%
\rightarrow }...\overset{u^{k}}{\rightarrow }\mu ^{k+1}=x(k+1)\overset{%
u^{k+1}}{\rightarrow }...
\end{equation*}%
To be remarked the appearance of the next state partial function $\forall
k\in \mathbf{N},\mathbf{B}^{n}\times U\ni (\mu ^{k},u^{k})\rightarrow \mu
^{k+1}\in \mathbf{B}^{n}.$ If $\exists k\in \mathbf{N}$ such that $%
u^{k}=u^{k+1}=...$ and $\mu ^{k+1}=\mu ^{k+2}=...,$ then the evolution may
be considered to be given by a finite sequence%
\begin{equation*}
\mu ^{0}=x(0)\overset{u^{0}}{\rightarrow }\mu ^{1}=x(1)\overset{u^{1}}{%
\rightarrow }...\overset{u^{k}}{\rightarrow }\mu ^{k+1}=x(k+1).
\end{equation*}
\end{remark}

\begin{example}
We get back to the system $f$ from the Example \ref{Exa63}, that fulfills
the non-anticipation property from Definition \ref{Def62}. We suppose that
the sequence $(t_{k})\in Seq$ satisfies $\forall k\in \mathbf{N},$%
\begin{equation*}
t_{2k+1}\geq t_{2k}+d_{r},\;t_{2k+2}\geq t_{2k+1}+d_{f}
\end{equation*}%
and let be the sequences $(u^{k})\in S,(\mu ^{k})\in \mathbf{B},$%
\begin{equation*}
u^{0}(t)=\chi _{\lbrack t_{0},\infty )}(t),u^{2}(t)=\chi _{\lbrack
t_{0},t_{1})}(t)\oplus \chi _{\lbrack t_{2},\infty )}(t),
\end{equation*}%
\begin{equation*}
u^{4}(t)=\chi _{\lbrack t_{0},t_{1})}(t)\oplus \chi _{\lbrack
t_{2},t_{3})}(t)\oplus \chi _{\lbrack t_{4},\infty )}(t),...
\end{equation*}%
\begin{equation*}
u^{1}(t)=\chi _{\lbrack t_{0},t_{1})}(t),u^{3}(t)=\chi _{\lbrack
t_{0},t_{1})}(t)\oplus \chi _{\lbrack t_{2},t_{3})}(t),...
\end{equation*}%
\begin{equation*}
\mu ^{0}=\mu ^{2}=\mu ^{4}=...=0,\mu ^{1}=\mu ^{3}=\mu ^{5}=...=1.
\end{equation*}%
For%
\begin{equation*}
u(t)=\chi _{\lbrack t_{0},t_{1})}(t)\oplus \chi _{\lbrack
t_{2},t_{3})}(t)\oplus \chi _{\lbrack t_{4},t_{5})}(t)\oplus ...
\end{equation*}%
we have%
\begin{equation*}
\forall x\in f(u^{0}),x(t)=x(t)\cdot \chi _{(t_{0},t_{0}+d_{r})}(t)\oplus
\chi _{\lbrack t_{0}+d_{r},\infty )}(t),
\end{equation*}%
\begin{equation*}
\forall x\in f(u^{1}),x(t)=x(t)\cdot \chi _{(t_{0},t_{0}+d_{r})}(t)\oplus
\chi _{\lbrack t_{0}+d_{r},t_{1}]}(t)\oplus x(t)\cdot \chi
_{(t_{1},t_{1}+d_{f})}(t),
\end{equation*}%
\begin{equation*}
\forall x\in f(u^{2}),x(t)=x(t)\cdot \chi _{(t_{0},t_{0}+d_{r})}(t)\oplus
\chi _{\lbrack t_{0}+d_{r},t_{1}]}(t)\oplus x(t)\cdot \chi
_{(t_{1},t_{1}+d_{f})}(t)\oplus 
\end{equation*}%
\begin{equation*}
\oplus x(t)\cdot \chi _{(t_{2},t_{2}+d_{r})}(t)\oplus \chi _{\lbrack
t_{2}+d_{r},\infty )}(t),
\end{equation*}%
\begin{equation*}
\forall x\in f(u^{3}),x(t)=x(t)\cdot \chi _{(t_{0},t_{0}+d_{r})}(t)\oplus
\chi _{\lbrack t_{0}+d_{r},t_{1}]}(t)\oplus x(t)\cdot \chi
_{(t_{1},t_{1}+d_{f})}(t)\oplus 
\end{equation*}%
\begin{equation*}
\oplus x(t)\cdot \chi _{(t_{2},t_{2}+d_{r})}(t)\oplus \chi _{\lbrack
t_{2}+d_{r},t_{3}]}(t)\oplus x(t)\cdot \chi _{(t_{3},t_{3}+d_{f})}(t)
\end{equation*}%
\begin{equation*}
...
\end{equation*}%
We conclude that $u$ is a fundamental mode of $f$:%
\begin{equation*}
\forall x\in f(u^{0}),x_{|(-\infty ,t_{0})}=0\;and\;x_{|[t_{1},\infty )}=1,
\end{equation*}%
\begin{equation*}
u_{|(-\infty ,t_{1})}=u_{|(-\infty ,t_{1})}^{0},\;u_{|(-\infty
,t_{2})}=u_{|(-\infty ,t_{2})}^{1},\;u_{|(-\infty ,t_{3})}=u_{|(-\infty
,t_{3})}^{2},...
\end{equation*}%
\begin{equation*}
\forall x\in f(u^{1}),x_{|[t_{2},\infty )}=0,\;\forall x\in
f(u^{2}),x_{|[t_{3},\infty )}=1,\;\forall x\in f(u^{3}),x_{|[t_{4},\infty
)}=0,...
\end{equation*}
\end{example}

\section{Accessibility vs fundamental mode}

\begin{theorem}
\label{The183}Let be the non-anticipatory system (Definition \ref{Def62}) $%
f:U\rightarrow P^{\ast }(S^{(n)}),U\in P^{\ast }(S^{(m)})$ and we suppose
that the following requirements are fulfilled:

$a)$ for any $(t_{k})\in Seq$ and any $(u^{k})\in U,$ we have $u^{0}\cdot
\chi _{(-\infty ,t_{0})}\oplus u^{1}\cdot \chi _{\lbrack t_{0},t_{1})}\oplus
u^{2}\cdot \chi _{\lbrack t_{1},t_{2})}\oplus ...\in U;$

$b)$ $f$ has race-free initial states and bounded initial time, i.e.%
\begin{equation*}
\forall u\in U,\exists \mu \in \mathbf{B}^{n},\exists t\in \mathbf{R}%
,\forall x\in f(u),x_{|(-\infty ,t)}=\mu ;
\end{equation*}

$c)$\ any vector from $\mathbf{B}^{n}$ is the common final value of the
states under an input having arbitrary initial segment%
\begin{equation*}
\forall \mu \in \mathbf{B}^{n},\forall u\in U,\forall t\in \mathbf{R}%
,\exists v\in U,\exists t^{\prime }>t,
\end{equation*}%
\begin{equation*}
u_{|(-\infty ,t)}=v_{|(-\infty ,t)}\;and\;\forall y\in f(v),y_{|[t^{\prime
},\infty )}=\mu .
\end{equation*}%
Then there is some $\mu ^{0}\in \mathbf{B}^{n}$ such that for any sequence $%
\mu ^{k}\in \mathbf{B}^{n},k\geq 1$ of binary vectors, there are the
sequences $(t_{k})\in Seq,$ $u^{k}\in U,k\in \mathbf{N}$ and an input $%
\widetilde{u}\in U$ such that%
\begin{equation*}
\forall x\in f(u^{0}),x_{|(-\infty ,t_{0})}=\mu ^{0}\;and\;x_{|[t_{1},\infty
)}=\mu ^{1},
\end{equation*}%
\begin{equation*}
\widetilde{u}_{|(-\infty ,t_{1})}=u_{|(-\infty ,t_{1})}^{0},\;\widetilde{u}%
_{|(-\infty ,t_{2})}=u_{|(-\infty ,t_{2})}^{1},\;\widetilde{u}_{|(-\infty
,t_{3})}=u_{|(-\infty ,t_{3})}^{2},...
\end{equation*}%
\begin{equation*}
\forall x\in f(u^{1}),x_{|[t_{2},\infty )}=\mu ^{2},\;\forall x\in
f(u^{2}),x_{|[t_{3},\infty )}=\mu ^{3},\;\forall x\in
f(u^{3}),x_{|[t_{4},\infty )}=\mu ^{4},...
\end{equation*}
\end{theorem}

\begin{proof}
Let $v^{0}\in U$ be an arbitrary input. From b) we get the existence of $\mu
^{0}\in \mathbf{B}^{n}$ and $t_{0}\in \mathbf{R}$ depending on $v^{0},$ such
that%
\begin{equation}
\forall x\in f(v^{0}),x_{|(-\infty ,t_{0})}=\mu ^{0}.  \label{fm_1}
\end{equation}%
Let us fix the sequence $\mu ^{k}\in \mathbf{B}^{n},k\geq 1$ and an
arbitrary number $\delta >0.$ At this moment the property c) implies the
existence of $u^{0}\in U$ and $t_{1}>t_{0}+\delta $ such that%
\begin{equation*}
v_{|(-\infty ,t_{0})}^{0}=u_{|(-\infty ,t_{0})}^{0}\;and\;\forall x\in
f(u^{0}),x_{|[t_{1},\infty )}=\mu ^{1},
\end{equation*}%
of $u^{1}\in U$ and $t_{2}>t_{1}+\delta $ such that%
\begin{equation*}
u_{|(-\infty ,t_{1})}^{0}=u_{|(-\infty ,t_{1})}^{1}\;and\;\forall x\in
f(u^{1}),x_{|[t_{2},\infty )}=\mu ^{2},
\end{equation*}%
of $u^{2}\in U$ and $t_{3}>t_{2}+\delta $ such that%
\begin{equation*}
u_{|(-\infty ,t_{2})}^{1}=u_{|(-\infty ,t_{2})}^{2}\;and\;\forall x\in
f(u^{2}),x_{|[t_{3},\infty )}=\mu ^{3},
\end{equation*}%
\begin{equation*}
...
\end{equation*}

The way that $(t_{k})$ was constructed guarantees the fact that this
sequence belongs to $Seq$. Thus, by a), the input $\widetilde{u}$ defined as%
\begin{equation*}
\widetilde{u}=u^{0}\cdot \chi _{(-\infty ,t_{1})}\oplus u^{1}\cdot \chi
_{\lbrack t_{1},t_{2})}\oplus u^{2}\cdot \chi _{\lbrack t_{2},t_{3})}\oplus
...
\end{equation*}%
belongs to $U$. We have%
\begin{equation*}
\widetilde{u}_{|(-\infty ,t_{1})}=u_{|(-\infty ,t_{1})}^{0},\;\widetilde{u}%
_{|(-\infty ,t_{2})}=u_{|(-\infty ,t_{2})}^{1},\;\widetilde{u}_{|(-\infty
,t_{3})}=u_{|(-\infty ,t_{3})}^{2},...
\end{equation*}
\end{proof}

\begin{remark}
The request b) of the theorem, that the initial states $\mu $ are race-free
and the initial time $t$ is bounded shows the order of the four quantifiers $%
\forall u,\exists \mu ,\exists t,\forall x,$ that is the two existential
quantifiers are in the middle (the total number of possibilities is $3\times
3=9$).

The key request in the hypothesis of the theorem is however that of
controllability and accessibility from item c). We make the terminological
remark that due to the frequent confusion that exists in the literature
generated by the concepts of controllability and accessibility, we prefer to
call all such requests 'accessibility'.

The theorem states that, in certain conditions (which are fulfilled by the
system $f$ from Example \ref{Exa63} and by many other systems), the initial
state $\mu ^{0}$ exists such that for any sequence $\mu ^{k}\in \mathbf{B}%
^{n},k\geq 1$, the fundamental mode $\widetilde{u}\in U$ exists, making $f(%
\widetilde{u})$ access the values $\mu ^{0},\mu ^{1},\mu ^{2},...$
synchronously, in this order.
\end{remark}

\end{document}